\documentclass[sn-mathphys,Numbered,iicol]{sn-jnl}% Default with double column layout

\usepackage{graphicx}%
\usepackage{multirow}%
\usepackage{amsmath,amssymb,amsfonts}%
\usepackage{amsthm}%
\usepackage{mathrsfs}%
\usepackage[title]{appendix}%
\usepackage{xcolor}%
\usepackage{textcomp}%
\usepackage{manyfoot}%
\usepackage{booktabs}%
\usepackage{algorithm}%
\usepackage{algorithmicx}%
\usepackage{algpseudocode}%
\usepackage{listings}%
\newcommand{\ha}{\mbox{\small$\frac{1}{2}$}}
\newcommand{\lab}[1]{\label{#1}}
\newcommand{\re}[1]{(\ref{#1})}
\newcommand{\nn}{\nonumber}
\newcommand{ \Bf }[1]{\boldsymbol{#1}}
\newcommand{\s}[1]{\mathsf{#1}}
\newcommand{\sOm}{\mathsf{\Omega}}
\newcommand{\BOm}{\boldsymbol{\Omega}}
\newcommand{\Bom}{\boldsymbol{\omega}}
\newcommand{\D}[2]{{\rm d}^{#1}{#2}\,}
\newcommand{\vl}{\,\rule[-0.3ex]{0.5pt}{2.0ex}\,}
\newcommand{\vdl}{\,\rule[-0.3ex]{0.5pt}{2.0ex}\,\rule[-0.3ex]{0.5pt}{2.0ex}\,}
\begin{document}

\title[Asymptotical dynamics of askew-polarized spinning top]%
{Asymptotical dynamics of askew-polarized spinning top under the radiation reaction torque}

%%=============================================================%%
%% Prefix   -> \pfx{Dr}
%% GivenName    -> \fnm{Joergen W.}
%% Particle -> \spfx{van der} -> surname prefix
%% FamilyName   -> \sur{Ploeg}
%% Suffix   -> \sfx{IV}
%% NatureName   -> \tanm{Poet Laureate} -> Title after name
%% Degrees  -> \dgr{MSc, PhD}
%% \author*[1,2]{\pfx{Dr} \fnm{Joergen W.} \spfx{van der} \sur{Ploeg} \sfx{IV} \tanm{Poet Laureate}
%%                 \dgr{MSc, PhD}}\email{iauthor@gmail.com}
%%=============================================================%%

\author*[1]{\fnm{Askold}
\sur{Duviryak}}\email{duviryak@icmp.lviv.ua}

\affil*[1]{\orgdiv{Department for Computer Simulations of
Many-Particle Systems}, \orgname{Institute for Condensed Matter
Physics of NAS of Ukraine}, \orgaddress{\street{1~Svientsitskii
Street}, \city{Lviv}, \postcode{79011}, %\state{},
\country{Ukraine}}, \\ \url{http://orcid.org/0000-0003-0786-4664}}

%%==================================%%
%% sample for unstructured abstract %%
%%==================================%%

\abstract{Rotary dynamics of polarized composite particles as dipole
rigid bodies is described by the Euler equations singularly
perturbed by the radiation reaction torque. The Schott term is
accounted, and the reduction procedure lowering higher derivatives
is applied. Asymptotic methods of nonlinear mechanics are used to
analyze the rotary dynamics of askew-polarized spinning top.
Numerical estimates are relevant to the hypothetical
DAST-nanocrystals that might possess a huge dipole moment. }

\keywords{spinning top, radiation reaction, Schott term}

\pacs[MSC Classification]{34D05, 34D15, 70E20}

\maketitle

%%%%%%%%%%%%%%%%%%%%%%%%%%%% SECTION 1 %%%%%%%%%%%%%%%%%%%%%%%%%%%%%%%%%

\section{Introduction}
\renewcommand{\theequation}{1.\arabic{equation}}
\setcounter{equation}{0}

Nowadays, nanoparticles in optical traps can be spun up to GHz
\cite{RDHD18,JYRLYZ21}. On the other hand, the dipole moment of some
artificially created nanoparticles, such as inorganic nanocrystals
CdSe and CdS \cite{S-K06} or cellulose nanocrystals \cite{F-PBL14}
reaches hundreds and thousands Debyes. Under such trends, the
effects of radiative spindown may soon become observable
experimentally \cite{Duv20a}, and their theoretical description is
timely.

It is shown in \cite{Duv20a,Duv22a} that for the description of
rotary dynamics of free polarized particles of nanometer size spun
up even to GHz the nonrelativistic rigid body mechanics combined
with the electrodynamics in the dipole approximation is a sufficient
theoretical base. Even so, at least two different expressions for
the radiation reaction torque acting on a system of constituent
charges are known in literature \cite{L-L87E,Jac99}. They are
derived in different ways and differ by the term which is dependent
on the third time derivative of the dipole moment of the system.
This term is analogous to the Schott term which, in turn, is
relevant to the radiation reaction power \cite{Sch12}. Usually, the
Schott-like term is neglected and both expressions for the radiation
reaction torque are regarded equivalent. Actually, this is not true,
as it follows from the discussion about the role of the Schott term
in the radiation reaction problem, recently raised in the literature
\cite{Gro11,Sin16c,Duv22a}.

In the present paper the importance and account of the Schott-like
term in the rotary dynamics of dipole composite particles are
substantiated. Besides, examples of polarized nanocrystals to which
this dynamics may concern are considered.

When regarding the rotary dynamics of a dipole composite particle as
a rigid body one arrives at the Euler equations perturbed by the
radiation reaction torque. Accounting the Schott term brings higher
derivatives into dynamics and makes the Euler equations singularly
perturbed possessing nonphysical solutions. Thus higher derivatives
must anyway be removed from the final description of the dynamics.
The recipe used in this work and proposed in the previous ones
\cite{Duv20a,Duv22a} consists in the reduction of higher derivatives
by means of the unperturbed Euler equations. The alternative, which
is consonant with the traditional viewpoint \cite{L-L87E,Jac99}, is
a simple neglect of the Schott term, but this may change crucially
the evolution of the particle \cite{Duv22a}.

In \cite{Duv20a,Duv22a} the composite particle was modeled as an
axially-symmetric spinning top with the dipole moment directed along
the symmetry axis. The symmetry in this example simplifies the Euler
equations and makes them integrable in both cases: with and without
the Schott term. Different dynamics yield different evolutions and
predict different far future asymptotical states. The model matches
to the aforementioned cellulose nanocrystals possessing 4400 Debyes
of the dipole moment \cite{F-PBL14}. It will be shown that this
value may appear insufficient to study experimentally the
radiation reaction spindown.

Here we develope the rotary dynamics of the askew-polarized
axially-symmetric spinning top. This case is relevant to the
hypothetical organic DAST-nanocrystals which are not synthesized yet
but predicted to possess millions Debyes of a permanent dipole
moment \cite{MNS03}. The axial symmetry in this problem is broken
due to incline of the dipole moment. Consequently, the rotary
dynamics in described by a complicated nonlinear set of Euler
equations which is not integrable. We combine approximation methods
such as linearization, asymptotic expansion, averaging and numerical
integration in order to study general features and details of the
radiative slowdown of the askew-polarized spinning top.

The paper is organized as follows. In the section 2 a balance
equation for the angular momentum of a system of radiating charges
is considered. Two different expressions for the radiation reaction
torque which differ by the time derivative Schott term are
discussed, and the corresponding Euler equations describing a rotary
motion of a polarized spinning top are presented. The section 3 is
devoted to the rotary dynamics of the symmetric spinning top with
arbitrarily inclined dipole moment. The Euler equation is split into
a coordinate components, and a partial exact solution is found. In
section 4 approximation methods of nonlinear mechanics are used to
study the asymptotic behavior of the askew-polarized spinning top.
In subsections 4.1--4.4 several asymptotical solutions are found
analytically, which are approved by numerical integration in
subsection 4.5. In section 5 three special cases of the polarized
spinning top are considered -- with longitudinal polarization
(subsection 5.1), with transverse polarization (subsection 5.2), and
the spherical spinning top. The application of aforementioned
results to the cellulose nanocrystals and to the hypothetical
DAST-nanocrystals is discussed in section 6. The final section 7
includes a summary and conclusions of the paper.

%%%%%%%%%%%%%%%%%%%%%%%%%%%% SECTION 2 %%%%%%%%%%%%%%%%%%%%%%%%%%%%%%%%%

\section{Equation of motion of a polarized spinning top}
\renewcommand{\theequation}{2.\arabic{equation}}
\setcounter{equation}{0}

Here we consider a composite particle consisting of point-like
charges $q$ with masses $m$ located at positions $ \Bf r$. We proceed
from the slow-motion balance equation \cite[section 75]{L-L87E}:
%
%           Equation 2.1
\begin{equation}\lab{2.1}
\dot{ \Bf L}=\frac2{3c^3}\,\Bf{\frak d}\times\dddot{ \Bf{\frak d}}
\end{equation}
for the angular momentum $ \Bf L=\sum m\, \Bf r\times \Bf v$, where
$ \Bf{\frak d}=\sum q \Bf r$ is the dipole moment of the system, and a
summation runs over constituent charges. The expression in r.-h.s.
is the torque of the Abraham-Lorentz radiation reaction forces
acting on constituents of the composite particle. The torque of
other forces is supposed to vanish.

This is the case when the composite particle is considered as a
rigid body, i.e., a spinning top. A rotational motion of an
arbitrary point $ \Bf r(t)$ of the top can be presented as follows:
$ \Bf r(t)=\s O(t)\Bf\rho$, where $\s O(t)\in\ $SO(3) is a rotation
matrix, and $\Bf\rho$ is a constant position of the point in the
proper reference frame of the top. We use a higher derivative rigid
body kinematics in eq. \re{2.1} and arrive at the Euler equation of
rotary motion of the spinning top \cite{Duv20a,Duv22a}:
%
%           Equation 2.2
\begin{eqnarray}\lab{2.2}
\s I\dot\BOm+\BOm\times\s I\BOm =\frac2{3c^3} \Bf d\times\{ \Bf d\times(\Omega^2\BOm-\ddot\BOm)&&\nn\\
{} +( \Bf d\cdot\dot\BOm)\BOm+2( \Bf d\cdot\BOm)\dot\BOm\};&&
\end{eqnarray}
here $\s I=\vdl I_{ij}\vdl$ ($i,j=1,2,3$) is the inertia tensor, $ \Bf d\equiv\s O^{\rm T} \Bf{\frak d}=\sum q\Bf\rho$ is a constant dipole
moment of the spinning top in the proper reference frame, the vector
$\BOm$, dual to the skew-symmetric matrix $\sOm\equiv\s O^{\rm
T}\dot{\s O}$, is the angular velocity of the spinning top in its
proper reference frame, and $\Omega\equiv\vl\BOm\vl $.

The Euler equation \re{2.2} contains in its r.-h.s. the 2nd
derivative $\ddot{\BOm}$ multiplied by a small parameter $d^2/c^3$,
i.e., this equation is a singularly perturbed one, and it possesses
redundant solutions which describe non-physical runaway spin up
\cite{Duv20a}. This difficulty can be avoided by reducing
higher-order derivatives in small perturbation terms by usage of the
free-motion Euler equation, i.e., the equation of a rotary motion of
a free spinning top, and its differential consequence:
%
%           Equation 2.3
\begin{subequations}\label{2.3}
\begin{eqnarray}
\dot\BOm&=&-\s I^{-1}(\BOm\times\s I\BOm),\lab{2.3a}\\
\ddot\BOm&=&-\s I^{-1}(\dot\BOm\times\s I\BOm+\BOm\times\s I\dot\BOm)\nn\\
&=&\s I^{-1}\{(\BOm\times(\BOm\times\s I\BOm)\nn\\
&&{}-(\s I\BOm)\times\s
I^{-1}(\BOm\times\s I\BOm)\}\lab{2.3b}.
\end{eqnarray}
\end{subequations}
General explicit form of the reduced Euler equation is cumbersome
and omitted here.

The angular momentum balance equation \re{2.1} is treated also in
another way \cite{L-L87E,Jac99}: the radiation reaction torque in
r.-h.s. of \re{2.1} is recast into two terms:
%
%           Equation 2.4
\begin{subequations}\label{2.4}
\begin{eqnarray}
\dot{ \Bf L}=-\frac2{3c^3}\,\dot{ \Bf{\frak d}}\times\ddot{ \Bf{\frak
d}}&&
\lab{2.4a}\\
&+&\frac{\D{}{}}{\D{}{t}}\frac2{3c^3}\, \Bf{\frak
d}\times\ddot{ \Bf{\frak d}}. \lab{2.4b}
\end{eqnarray}
\end{subequations}
The total time derivative term \re{2.4b} containing the 3rd
derivative of the dipole moment $\dddot{ \Bf{\frak d}}$ and referred
here to as the Schott-like term \cite{Duv22a} is usually regarded as
negligibly small if, at least, the system moves quasi-periodically
\cite[section 75]{L-L87E}. Then the use of the rigid body kinematics in
the equation \re{2.4a} leads to the equation
%
%           Equation 2.5
\begin{eqnarray}\lab{2.5}
\s I\dot{\BOm}+\BOm\times\s I\BOm &=&-\frac2{3c^3}\{( \Bf d\times\BOm)^2\BOm\nn\\
&&{}+( \Bf d\cdot(\BOm\times\dot{\BOm})) \Bf d\}
\end{eqnarray}
which will be referred to as the truncated Euler equation. In
contrast to \re{2.2}, this equation is not singularly perturbed, and
can be reduced to the normal form without the use of the free-motion
Euler equations \re{2.3}.

The neglect of the time derivative term \re{2.4b} is substantiated
by an alternative derivation of the balance equation in the form
\re{2.4a}. It is based on accounting the flux of the angular
momentum of the dipole radiation through the sphere embracing
constituent charges \cite[section 72]{L-L87E}. Actually, the result is
the hereditary balance equation:
%
%           Equation 2.6
\begin{eqnarray}
\dot{\Bf{L}}{\,\rule[-5pt]{0.5pt}{14pt}\,}_{t}=-\frac2{3c^3}\,\dot{\Bf{\frak
d}}\times\ddot{\Bf{\frak d}}{\,\rule[-5pt]{0.5pt}{14pt}\,}_{t-R/c} \lab{2.6}
\end{eqnarray}
which is the more precise, the larger is the radius $R$ of the
sphere. In the limit $R\to\infty$ it turns into the equations with
infinitely retarded time argument. This fact is usually not
emphasized, and the balance equation \re{2.6} might be treated
mistakenly as the instantaneous one \re{2.4a}. Dynamical
consequences of this confusion will be displayed in next sections.

%%%%%%%%%%%%%%%%%%%%%%%%%%%% SECTION 3 %%%%%%%%%%%%%%%%%%%%%%%%%%%%%%%%%

\section{Dynamics of an axially symmetric spinning top with inclined dipole moment}
\renewcommand{\theequation}{3.\arabic{equation}}
\setcounter{equation}{0}

Let us consider the case of the symmetric spinning top with the
dipole moment inclined to a symmetry axis by the angle $\chi$. In
the properly oriented reference frame, where the ort $ \Bf e_3$ is
directed along the symmetry axis of the top, and the dipole moment
$ \Bf d$ lies in the plane O$ \Bf e_1 \Bf e_3$, we have:
%
%           Equation 3.1-2
\begin{eqnarray}
I_{ij}&=&I_i\delta_{ij}, \quad I_2=I_1, \quad 0<I_3\le2I_1,~~~
\lab{3.1}\\
 \Bf d&=&\{d_1,0,d_3\},\quad
d_1=d\sin\chi,\nn\\
\quad d_3&=&d\cos\chi,\qquad 0\le\chi\le\pi/2 \lab{3.2}
\end{eqnarray}
(there is no summation over $i$); see figure \ref{fig1}.

We will consider parameters $I_1$, $I_3$, $d$, $\chi$ arbitrary.
Then the reduced equation of motion \re{2.2} splits into the
following ones:
%
%           Equation 3.3-6
%
{\small
\begin{eqnarray}
\dot\Omega_1-\delta\,\Omega_2\Omega_3&=&-\frac{2d_3}{3I_1c^2}\{
d_3\Omega_a^2\Omega_1- d_1\Omega_b^2\Omega_3 \},~~~~~
\lab{3.3}\\
\dot\Omega_2+\delta\,\Omega_1\Omega_3&=&-\frac{2}{3I_1c^2}\{d_1^2\Omega_c^2+d_3^2\Omega_a^2
\nn\\
&&\hspace{10ex}{}- 3d_1d_3\delta\,\Omega_1\Omega_3 \}\Omega_2,~~~~~
\lab{3.4}\\
\dot\Omega_3&=&-\frac{2d_1}{3I_3c^2}\{
d_1\Omega_b^2\Omega_3-d_3\Omega_a^2\Omega_1 \},~~~~~\lab{3.5}
\end{eqnarray}
}
where
\begin{eqnarray}
\Omega_a^2&\equiv&\Omega^2+\delta(\delta-2)\Omega_3^2,
\nn\\
\Omega_b^2&\equiv&\Omega^2+\delta(2\Omega_1^2{-}\Omega_2^2),
\nn\\
\Omega_c^2&\equiv&\Omega^2+\delta(\delta+1)\Omega_3^2,\nn
\end{eqnarray}
and the parameter
\begin{equation}\lab{3.6}
\delta\equiv1-I_3/I_1,\qquad -1\le\delta<1
\end{equation}
characterizes an elongacy of the inertia spheroid.

%---------------------- Figure 1 ----------------------------
\begin{figure}[t]
\begin{center}
\includegraphics[scale=0.5]{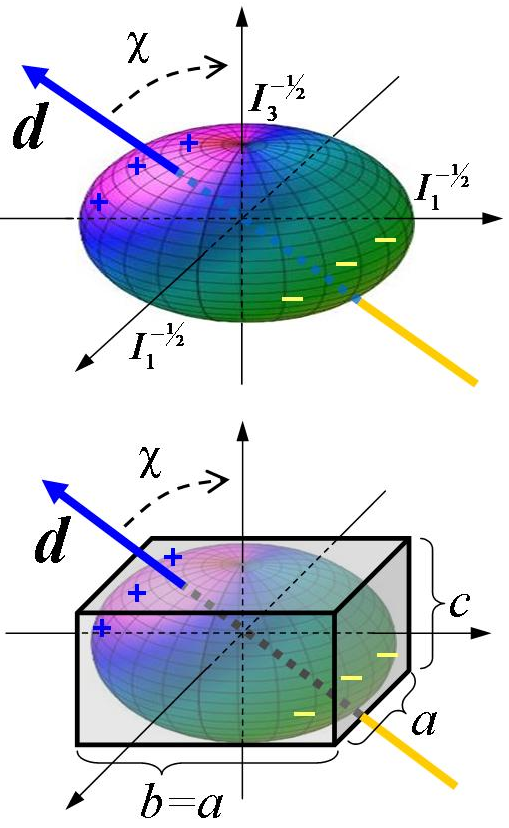}
%\vspace{-2ex}
\caption{The model of askew-polarized crystal particle (center). The
dipole moment vector $ \Bf d$ is inclined to the symmetry axis by
the angle $\chi$. The axially-symmetric inertia ellipsoid (i.e., the
spheroid) of the crystal (shown inside) is characterized by the
principal inertia moments $I_1=I_2$ and $I_3$ (top). The model
comprises rigid bodies of more irregular shape provided their
inertia ellipsoids are axially-symmetric (bottom). }\lab{fig1}
\includegraphics[scale=0.4]{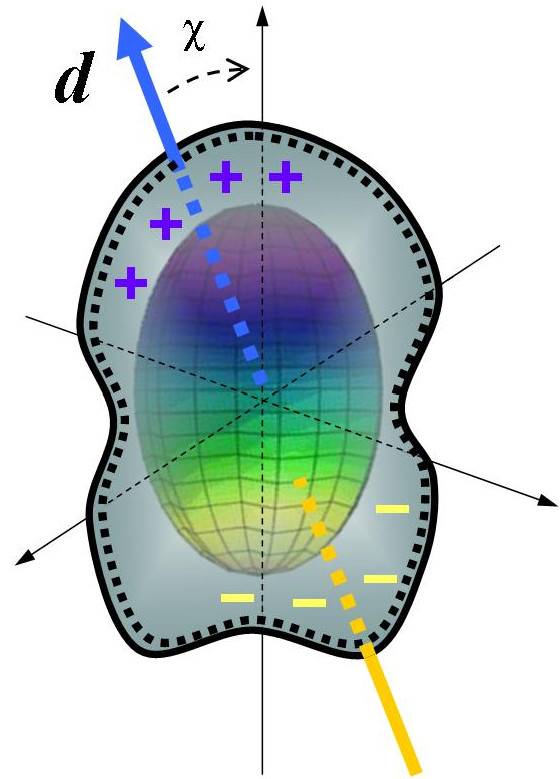}
\end{center}
\end{figure}
%------------------------------------------------------------

One can find easily a particular exact solution corresponding to
$\Omega_1=\Omega_3=0$. In this case the equations \re{3.3} and
\re{3.5} become identities while the equation \re{3.4} simplifies,
%
%           Equation 3.7
\begin{eqnarray}
\dot\Omega_2&=&-\tau_0\Omega_2^3, \lab{3.7}
\end{eqnarray}
where we introduced the time-scale parameter
%
%           Equation 3.8
\begin{equation}\lab{3.8}
\tau_0\equiv\frac{2d^2}{3I_1c^3},
\end{equation}
and possesses the particular solution
%
%           Equation 3.9
\begin{align}\lab{3.9}
&\Omega_2(t)=\Omega_{20}/F(t), \qquad \Omega_2(0)=\Omega_{20},\nn\\
\mbox{where}\quad &F(t)\equiv\sqrt{1+2\tau_0\Omega_{20}^2t}.
\end{align}
It describes a rotary braking of the spinning top which is most
intensive during the characteristic time
$T=1/(\tau_0\Omega_{20}^2)$. Then, at $t\gg T$, the asymptotic
regime comes, and the magnitude $\Omega_2$ of the angular velocity
ceases by the power law $\Omega_2\sim\pm1/\sqrt{2\tau_0t}$ which
does not depend on the initial values. Actually, the partial
solution \re{3.9} coincides with its asymptotics. This becomes
evident after redefining the time variable by a finite shift: $t\to
t'=t+T/2$.

General exact solution of the set \re{3.3}--\re{3.5} is unknown,
thus we apply to approximations.

%%%%%%%%%%%%%%%%%%%%%%%%%%%% SECTION 4 %%%%%%%%%%%%%%%%%%%%%%%%%%%%%%%%%

\section{Asymptotical analysis of askew-polarized spinning top dynamics}
\renewcommand{\theequation}{4.\arabic{equation}}
\setcounter{equation}{0}

Let us introduce dimensionless variables:
%
%           Equation 4.1
\begin{equation}\lab{4.1}
\tau\equiv t/\tau_0,\qquad \Bom\equiv\tau_0\BOm,
\end{equation}
where $\tau_0$ is defined in eq. \re{3.8}. In these terms the system
\re{3.3}--\re{3.5} takes the form:
%
%           Equation 4.2-4.4
\begin{eqnarray}
\dot\omega_1-\delta\,\omega_2\omega_3&=& - n_3^2\omega_a^2\omega_1+
n_1n_3\omega_b^2\omega_3,
\lab{4.2}\\
\dot\omega_2+\delta\,\omega_1\omega_3&=&-\{n_1^2\omega_c^2+n_3^2\omega_a^2\nn\\
&&{}- 3n_1n_3\delta\,\omega_1\omega_3 \}\omega_2,
\lab{4.3}\\
(1-\delta)\dot\omega_3&=&-n_1^2\omega_b^2\omega_3+n_1n_3\omega_a^2\omega_1,
\lab{4.4}
\end{eqnarray}
where $n_1\equiv\sin\chi$, $n_3\equiv\cos\chi$, $0<\chi<\pi/2$ (in
this section) and the dot ``$\dot{\phantom{\omega}}$'' denotes
differentiation with respect to $\tau$. Combining equations \re{4.2}
and  \re{4.4} yields the equation
%
%           Equation 4.5
\begin{eqnarray}
n_1\dot\omega_1+n_3(1-\delta)\dot\omega_3-n_1\delta\,\omega_2\omega_3&=&0
\lab{4.5}
\end{eqnarray}
which sometimes is convenient to use, instead of eq. \re{4.2}.

The dimensionless system \re{4.2}--\re{4.4} or the equivalent one
\re{4.3}--\re{4.5} represents two-parametrical family of dynamical
systems characterized by the elongacy parameter $\delta$ and the
inclination angle $\chi$. It is remarkable that the system does not
include any small perturbation parameter manifestly. On the other
hand, the time scale parameter $\tau_0$ is actually small. Looking
ahead let us note that even for strongly polarized DAST-nanocrystal
mentioned above $\tau_0\sim10^{-27}\,$c. Thus, the dimensionless
time $\tau$ is ``fast'', components of the dimensionless angular
velocity $ \Bf\omega$ are small and ``slow'', and r.-h.s of the set
\re{4.2}--\re{4.4} is small compared to separate terms in l.-h.s.
This is useful for approximations.

\subsection{Power-law monotonous asymptotics}

The only fixed point of the system \re{4.3}--\re{4.5} is $\Bom=0$.
Let us consider an asymptotical behavior of $\Bom$ at $\tau\to\infty$.
First of all, we examine the power-law asymptotical behavior and
represent the components $\omega_1$, $\omega_2$, $\omega_3$ as
principal terms of the asymptotic expansion power series,
%
%           Equation 4.6
\begin{eqnarray}\lab{4.6}
\omega_1&=&A\,\tau^\alpha[1+O(\tau^{-1})],\quad
\omega_2=B\,\tau^\beta[1+O(\tau^{-1})],\nn\\
\omega_3&=&C\,\tau^\gamma[1+O(\tau^{-1})],
\end{eqnarray}
similarly to the case of a single nonlinear differential equation
\cite{Bel53}. Substituting \re{4.6} into \re{4.5} yields the
equation:
%
%           Equation 4.7
\begin{eqnarray}\lab{4.7}
n_1\alpha A\tau^{\alpha-1} + n_3(1-\delta)\gamma C\tau^{\gamma-1}-
n_1\delta BC\tau^{\beta+\gamma}\nn\\
 =O(\tau^{\alpha-2}) +
O(\tau^{\gamma-2}) + O(\tau^{\beta+\gamma-1})~~~~~~~
\end{eqnarray}
in which all or at least two leading terms on the l.-h.s. must
compensate one another while the remaining term as well as all the
right-hand side of eq. \re{4.7} must be negligibly small compared to
the leading terms. The equations \re{4.3} and \re{4.4} are treated
in similar manner. In such a way we derive the algebraic set of
equations for the constants $A$, $B$, $C$, $\alpha$, $\beta$,
$\gamma$. Since there is not known in advance which terms are
leading and which are minor, one should examine all combinations. As
a result, we arrive at two self-consistent possibilities.

The first one,
%
%           Equation 4.8
\begin{eqnarray}
A=C=0,\quad B=\pm1/\sqrt2,\quad \beta=-1/2, \lab{4.8}
\end{eqnarray}
restores the exact partial solution \re{3.9}. The second solution,
eqs. \re{4.6} with
%
%           Equation 4.9
\begin{eqnarray}
\gamma&=&-1/2,\quad\beta=-1,\quad\alpha=-3/2,
\nn\\
C&=&\pm\sqrt{\frac{1-\delta}{2n_1^2}},\quad
B=-\frac{n_3(1-\delta)}{2n_1\delta},\nn\\
A&=&\frac{B}{\delta
C}\{1-[(1-\delta)^2+3n_1^2\delta]C^2\},~~~ \lab{4.9}
\end{eqnarray}
is not exact but only asymptotical. All components of $\Bom$ reveal
a monotonous decrease at $\tau\to\infty$ so that
$\omega=\vl\Bom\vl\sim1/\sqrt{\tau}$ and $\Bom\to0$. Amplitudes of this
asymptotics are rigidly fixed by parameters of the system. Thus it
is plausible that some unique selection or rather narrow set of
initial conditions leads to this asymptotics which we will label as
the asymptotical solution type ``1''. There exist other
asymptotical solutions analyzed in next subsections and approved by
numerical calculations in subsection 4.5.

\subsection{Spiral asymptotics}

One should expect that, if $\omega\ll1$, the dynamical system
\re{4.2}--\re{4.4} behaves mainly as a weakly perturbed spinning
top, i.e., its angular velocity precess around the symmetry axis
with slowly varying speed and magnitude. Thus we split the angular
velocity vector $ \Bf\omega$ into longitudinal and transverse
components, and the latter present as follows
%
%           Equation 4.10
\begin{equation}
\omega_1=\omega_\bot\cos\varphi, \qquad
\omega_2=\omega_\bot\sin\varphi,
\end{equation}
where the angle $\varphi$ determines a direction of the transverse
vector $ \Bf\omega_\bot=\{\omega_1,\omega_2,0\}$, and
$\omega_\bot=\vl\Bf\omega_\bot\vl$. In these terms the equations
\re{4.2}--\re{4.4} take the form:\vspace{-5ex}

\begin{strip}
%
%           Equation 4.11
\begin{eqnarray}
\dot\omega_\bot&=&-n_3^2\omega_a^2\omega_\bot -\ha
n_1^2\omega_c^2\omega_\bot(1-\cos2\varphi) +
n_1n_3[\omega^2+2\delta\omega_\bot^2]\omega_3\cos\varphi,
\nn\\
(1-\delta)\dot\omega_3&=&-n_1^2[\omega^2+\ha\delta\omega_\bot^2(1+3\cos2\varphi)]\omega_3
+ n_1n_3\omega_a^2\omega_\bot\cos\varphi,
\nn\\
\omega_\bot(\dot\varphi&+&\delta\omega_3)=-\ha
n_1^2\omega_c^2\omega_\bot\sin2\varphi -
n_1n_3[\omega^2-\delta\omega_\bot^2]\omega_3\sin\varphi. \lab{4.11}
\end{eqnarray}
\end{strip}

\noindent
It follows that components $\omega_\bot$ and $\omega_3$
are slow variables while the angle $\varphi$ is a fast variable.
Thus one can apply the averaging method of nonlinear mechanics
\cite{M-D97}. Averaging the r.-h.s. of the equations \re{4.11} over
the fast variable $\varphi$ yields the system for the averaged
variables $\bar\omega_\bot$, $\bar\omega_3$, $\bar\varphi$:
%
%           Equation 4.12
\begin{eqnarray}
\dot{\bar\omega}_\bot&=&-\{n_3^2[\bar\omega_\bot^2+(1-\delta)^2\bar\omega_3^2]\nn\\
&&{}+ \ha
n_1^2[\bar\omega_\bot^2+(1+\delta+\delta^2)\bar\omega_3^2]\}\bar\omega_\bot,
\nn\\
(1-\delta)\dot{\bar\omega}_3&=&-\ha
n_1^2[(2+\delta)\bar\omega_\bot^2+2\bar\omega_3^2]\bar\omega_3,
\nn\\
\dot{\bar\varphi}&=&-\delta\bar\omega_3. \lab{4.12}
\end{eqnarray}
This system is exactly solvable as it is similar to that presented
in \cite[eqs.\,4.9-11]{Duv20a}. But here we are interested in
asymptotics.

Again, ${\bar\omega}_\bot$ and ${\bar\omega}_3$ are slow variables
while $\bar\varphi$ is fast. We put
%
%           Equation 4.13
\begin{eqnarray}\lab{4.13}
\bar\omega_\bot&=&A\,\tau^\alpha[1+O(\tau^{-1})],\quad
\bar\omega_3=B\,\tau^\beta[1+O(\tau^{-1})],\nn\\
\bar\varphi&=&C\,\tau^\gamma[1+O(\tau^{-1})],
\end{eqnarray}
and substitute into eqs. \re{4.12}. The values of constants
$A,B,C,\alpha,\beta,\gamma$ to be found depend on the parameters
$\delta$ and $\xi\equiv\tan^2\chi$. There are three different cases
marked in follows the asymptotics type ``1'' (in subsection
4.1):

\begin{strip}
%
%           Equation 4.14-4.19
\begin{eqnarray}
2.&&
\alpha=-\frac{(1-\delta^3)}4-\frac{(1-\delta)^3}{2\xi}<\beta=-\frac12,\qquad
B^2=\frac{1-\delta}{2n_1^2}
\lab{4.14}\\
&\mbox{if}&\qquad 0<\xi<2\frac{(1-\delta)^3}{1+\delta^3}.
\lab{4.15}\\
3.&& \alpha=\beta=-\frac12,\qquad
A^2=\frac{(1+\delta^3)\xi-2(1-\delta)^3}{\delta n_1^2[2(3-\delta^2)-(3+3\delta+\delta^2)\xi]},\nn\\
&&\hspace{17ex}B^2=\frac{2(1-\delta)-(1+2\delta)\xi}{\delta
n_1^2[2(3-\delta^2)-(3+3\delta+\delta^2)\xi]}
\lab{4.16}\\
&\mbox{if}&\qquad
\frac{2(1-\delta)^3}{1+\delta^3}\le\xi\quad\mbox{and}\quad\xi\le\frac{2(1-\delta)}{1+2\delta},\quad
\delta>-\frac12.
\lab{4.17}\\
4.&&
\beta=-\frac{(2+\delta)\xi}{2(1-\delta)(\xi+2)}<\alpha=-\frac12,\qquad
A^2=\frac1{2-\xi}
\lab{4.18}\\
&\mbox{if}&\qquad \xi>\frac{2(1-\delta)}{1+2\delta},\quad
\delta>-\frac12,\quad\mbox{and}\quad
\xi<\frac{4(1-\delta)}{3\delta},\quad \delta>0. \lab{4.19}
\end{eqnarray}
\end{strip}
%

%
%---------------------- Figure 2 ----------------------------
\begin{figure}
\begin{center}
\includegraphics[scale=0.28]{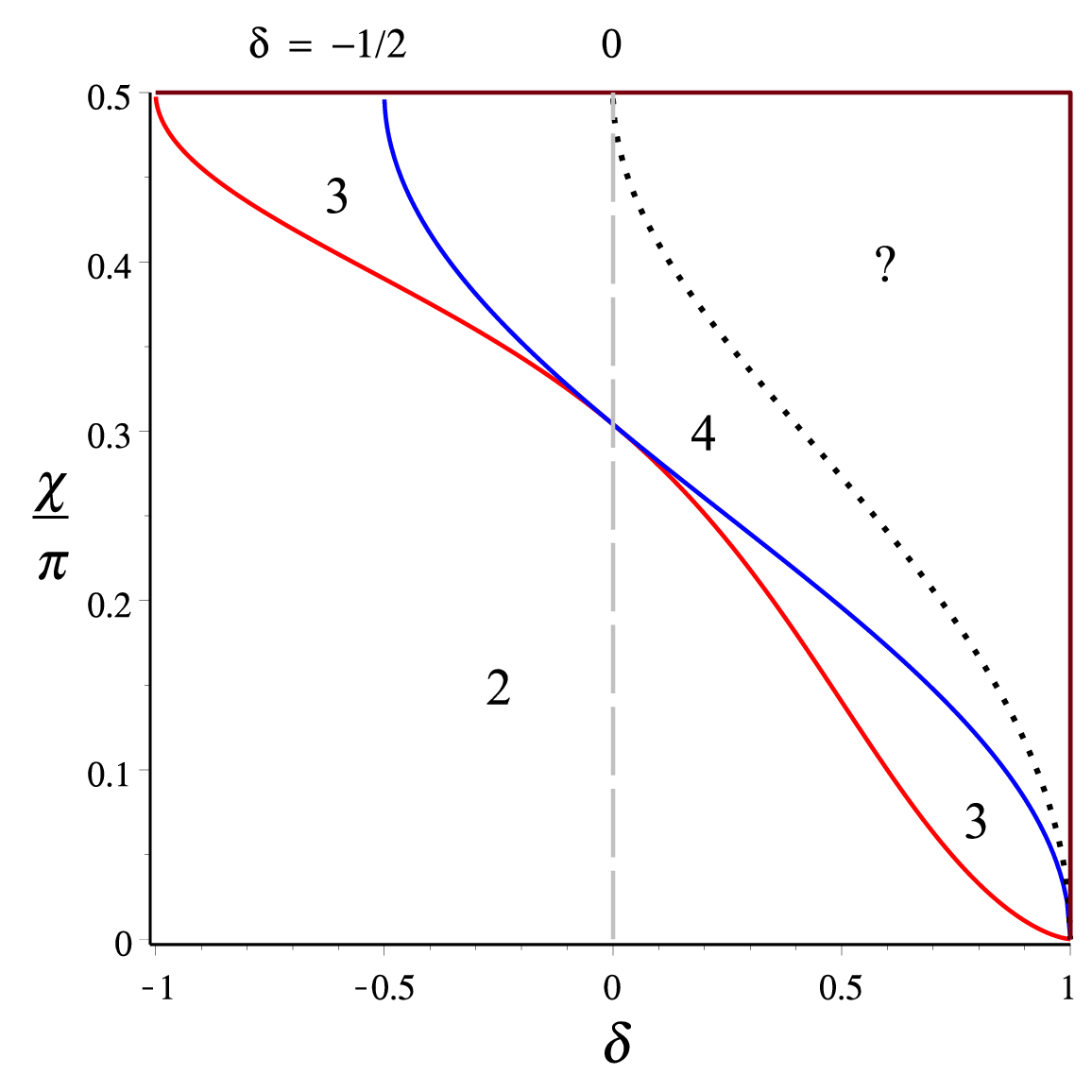}
\caption{Domains of the $\delta$--$\chi$ parameter plane in which
the spiral asymptotical solutions ``2--4'' exist. In the domain
``?'' the ``fast'' variable $\bar\varphi$ ``stops'':
$\bar\varphi\sim\tau^\gamma$, $\gamma<0$, so that the averaging
makes no sense.} \lab{fig2}
\end{center}
\end{figure}
%------------------------------------------------------------

\noindent
The magnitude $A$ in the case ``2'' and $B$ in the case
``4'' remain undetermined within the approximation used.
Nevertheless, in all three cases we have $\omega=\vl\Bom\vl
\sim1/\sqrt{\tau}$ at $\tau\to\infty$. Domains of the
$\delta$--$\chi$ parameter plane  in which these spiral asymptotical
solutions ``2--4'' exist are outlined in figure \ref{fig2}. In
particular, the domain ``3'' (eq. \re{4.17}) of the solution
\re{4.13}, \re{4.16} consists of two components conjunct by an
isthmus at the point $\delta=0$, $\chi_0=\arctan\sqrt{2}=0.304\pi$.

The averaging of the system \re{4.11} over the variable $\varphi$
makes a sense if $\gamma>0$. This condition violates if
%
%           Equation 4.20
\begin{equation}\lab{4.20}
\xi>\frac{4(1-\delta)}{3\delta},\quad \delta>0.
\end{equation}
This means that the assumption about a monotonous decrease of
variables $\omega_\bot$, $\omega_3$ and a monotonous increase of
$\varphi$ is not true in the domain \re{4.20} marked in figure
\ref{fig2} as ``?''. Thus one should look in this domain for
an asymptotic solution different from spiral ones.

\subsection{Oscillating solution on a time segment}

Coming back to the exact particular solution \re{3.9} of the system
\re{3.3}--\re{3.5} and redefining the time variable $t$ by a finite
shift, we have $\Omega_1=\Omega_3=0$,
$\Omega_2=\pm1/\sqrt{2\tau_0t}$; see the end of section 3. In terms
of dimensionless variables \re{4.1} this solution reads:
$\omega_1=\omega_3=0$, $\omega_2=\pm1/\sqrt{2\tau}$. Let us consider
a small perturbation to this solution:
%
%           Equation 4.21-22
\begin{eqnarray}
h\equiv \omega_2-f,\qquad \mbox{where}\quad f\equiv\pm1/\sqrt{2\tau}
\lab{4.21}
\end{eqnarray}
and
\begin{eqnarray}
\vl\omega_1\vl, \vl\omega_3\vl, \vl h\vl \ll
\vl f\vl=1/\sqrt{2\tau}. \lab{4.22}
\end{eqnarray}
Introducing a new small variable (instead of $\omega_1$):
%
%           Equation 4.23
\begin{equation}\lab{4.23}
\omega_\angle\equiv\omega_1+\frac{n_3}{n_1}(1-\delta)\omega_3
\end{equation}
and keeping in eqs. \re{4.5}, \re{4.4} terms up to linear ones in
small variables, one obtains:
%
%           Equation 4.24-4.25
\begin{eqnarray}
\dot\omega_\angle&=&\delta f\omega_3,
\lab{4.24}\\
\dot{\omega}_3&=&f^2\left[\frac{n_1n_3}{1-\delta}\omega_\angle-\omega_3\right].
\lab{4.25}
\end{eqnarray}
Eliminating $\omega_3$ yields the 2nd-order equation:
%
%           Equation 4.26
\begin{equation}\lab{4.26}
\ddot\omega_\angle+\frac1\tau\dot\omega_\angle\mp\frac{n_1n_3}{2\sqrt{2}(1-\delta)}\omega_\angle=0
\end{equation}
the solution of which can be expressed in terms of Bessel functions.
In order to have a decreasing asymptotics $\Bom\to0$ at
$\tau\to\infty$, we chose $f(\tau)\equiv-1/\sqrt{2\tau}$ and put
$\delta>0$ (when in the domain ``?'', eq. \re{4.20}). Using,
besides, eq. \re{4.24} one gets:
%
%           Equation
\begin{eqnarray*}
\omega_\angle(\tau)&=&\frac{4\delta}{\sqrt2a}\left\{AJ_0(a\tau^{1/4})+BN_0(a\tau^{1/4})\right\},
\\
\omega_3(\tau)&=&\tau^{-1/4}\left\{AJ_1(a\tau^{1/4})+BN_1(a\tau^{1/4})\right\},
\end{eqnarray*}
where
$\displaystyle{a=2^{5/4}\sqrt{\frac{n_1n_3\delta}{1-\delta}}}$. In
asymptotics $\tau\to0$
%
%           Equation 4.27
\begin{eqnarray}
\omega_1(\tau)&\approx&\omega_\angle(\tau)=\frac{4\delta}{\sqrt2a}C\tau^{-1/8}\cos(a\tau^{1/4}),
\nn\\
\omega_3(\tau)&=&C\tau^{-3/8}\sin(a\tau^{1/4}), \lab{4.27}
\end{eqnarray}
where $A$, $B$ and $C$ are arbitrary constants.

Let us consider the correction $h=\omega_2-f$. Substituting the
asymptotics \re{4.27} into eq. \re{4.3}, keeping leading terms in
$1/\tau$ and up to linear terms in the small variable $h$ we arrive
at the equation
%
%           Equation 4.28
\begin{equation}\lab{4.28}
\dot h = -\omega_1^2h-\delta\omega_1\omega_3
\end{equation}
which can be solved exactly. Missing details, we represent the
asymptotics of the solution:
%
%           Equation 4.29
\begin{equation}\lab{4.29}
h(\tau)\sim-\frac{a}{2\sqrt2\tau^{1/4}}\sin(2a\tau^{1/4}).
\end{equation}
The corrections \re{4.27} and \re{4.29} decrease more slowly than
the leading term $\vl\omega_2\vl=1/\sqrt{2\tau}$. Thus the approximated
solution \re{4.21}, \re{4.27}, \re{4.29} is not the asymptotics but
makes the sense at a finite time segment up to the moment when the
conditions \re{4.22} violate.

\subsection{Oscillating asymptotics}

The previous oscillating solution on a finite time segment suggests
a structure of a possible oscillating asymptotical solution. We
suppose that both solutions reveal similar oscillating behavior, but
components of the dimensionless angular velocity satisfy
asymptotically the relations:
$
\vl\omega_1\vl\sim\vl\omega_2\vl\gg\vl\omega_3\vl.
$
Taking these relations into account and using the substitution
\re{4.23} turns the system \re{4.3}--\re{4.5} into the exact
equation:
%
%           Equation 4.30
\begin{equation}\lab{4.30}
\dot\omega_\angle=\delta\omega_2\omega_3
\end{equation}
and the linearized ones:
%
%           Equation 4.31-4.32
\begin{align}
\dot{\omega}_2\approx&-[\omega_\angle^2+\omega_2^2]\omega_2
-\left\{\delta-\left[2(1-\delta)n_3/n_1\right.\right.&\nn\\
&{}+\left.\left.3n_1n_3\delta\right]\omega_2\right\}\omega_\angle\omega_3,&
\lab{4.31}\\
\dot{\omega}_3\approx&-\frac{n_1n_3}{1-\delta}[\omega_\angle^2+\omega_2^2]\omega_\angle\nn\\
&{}-\left[\left(n_1^2\frac{1+2\delta}{1-\delta}+3n_3^2\right)\omega_\angle^2+\omega_2^2\right]\omega_3.
\lab{4.32}
\end{align}
Then we put:
%
%           Equation 4.33-34
\begin{eqnarray}
\omega_\angle&=&f\cos\varphi, \quad \omega_2=g+h, \nn\\
\omega_3&=&v\sin\varphi+w\cos\varphi
\lab{4.33}\\
\mbox{where}\quad f, g &\gg& h, v, w \lab{4.34}
\end{eqnarray}
and $\varphi$ are functions of $\tau$ to be found.

Substituting \re{4.33} into \re{4.30} and accounting  \re{4.34} to
neglect small terms yields the equation:
%
%           Equation
\begin{equation*}
[\dot f-\delta gw]\cos\varphi+[f\dot\varphi+\delta gv]\sin\varphi=0
\end{equation*}
which being multiplied by $\cos\varphi$ or  $\sin\varphi$ and
averaged over $\varphi$ splits into the pair of equations:
%
%           Equation 4.35-36
\begin{eqnarray}
\dot f=\delta gw,
\lab{4.35}\\
\dot\varphi=-\delta\frac gfv. \lab{4.36}
\end{eqnarray}
When treating eq. \re{4.32} in similar manner one arrives at the
following pair of equations:
%
%           Equation 4.37-4.38
\begin{align}
\dot v=-&\delta\frac gfwv\nn\\
-&\left[\frac14\left(n_1^2\frac{1+2\delta}{1-\delta}+3n_3^2\right)f^2+g^2\right]v,
\lab{4.37}\\
\dot w=\delta&\frac gfv^2+\frac{n_1n_3}{1-\delta}\left[\frac34f^2+g^2\right]f\nn\\
-&\left[\frac34\left(n_1^2\frac{1+2\delta}{1-\delta}+3n_3^2\right)f^2+g^2\right]w.
\lab{4.38}
\end{align}
Defining $g=\bar\omega_2$ splits \re{4.31} into the pair:
%
%           Equation 4.39-4.40
\begin{align}
\dot g=-&\left[\frac12f^2+g^2\right]g,
\lab{4.39}\\
\dot h=-&\left[f^2\cos^2\varphi+3g^2\right]h\nn\\
{}-&\left\{\delta-\left[2\frac{n_3}{n_1}(1-\delta)+3n_1n_3\delta\right]g\right\}\times\nn\\
\times\,&f\cos\varphi\,\{v\sin\varphi+w\cos\varphi\}. \lab{4.40}
\end{align}
We look for asymptotical solutions of eqs. \re{4.35}--\re{4.39}:
%
%           Equation 4.41
\begin{align}\lab{4.41}
f&=A\,\tau^\alpha[1+O(\tau^{-1})],\quad
g=B\,\tau^\beta[1+O(\tau^{-1})],\nn\\
\varphi&=C\,\tau^\gamma[1+O(\tau^{-1})],\quad
v=M\,\tau^\mu[1+O(\tau^{-1})],\nn\\
w&=N\,\tau^\nu[1+O(\tau^{-1})]
\end{align}
and find the following values of parameters:
%
%           Equation 4.42
\begin{align}
\alpha&=\beta=-1/2, \quad \gamma=1/4,\nn\\
\mu&=-3/4,\quad \nu=-1,
\nn\\
A^2&=\frac{3(1+\xi)(1-\delta)}{1-\delta-(1-4\delta)\xi},\quad
&B=-\sqrt{\frac{1-A^2}2}\nn\\
C&=-4\delta\frac BA M,\quad &N=-\frac{A}{2\delta B},\nn\\
M^2&=\frac{n_1n_3A^2[3A^2+4B^2]}{4\delta(1-\delta)\vl B\vl},
\lab{4.42}
\end{align}
provided
%
%           Equation 4.43
\begin{eqnarray}
\xi&>&\frac{2(1-\delta)}{7\delta-6},\qquad \delta>4/7. \lab{4.43}
\end{eqnarray}
Then we search the asymptotics for $h$ from eq. \re{4.40} as
%
%           Equation 4.44
\begin{equation}\lab{4.44}
h\sim K\tau^\varkappa\cos2\varphi
\end{equation}
and find
%
%           Equation 4.45
\begin{equation}\lab{4.45}
\varkappa=-1/2, \qquad K=A^2/(4B).
\end{equation}

We see that $\vl h\vl\ll\vl g\vl$ provided $\vl K\vl\ll\vl B\vl$ which is possible if
the first inequality in \re{4.43} is strong (i.e., ``$\gg$'' instead
of ``$>$'').

\subsection{Numerical solutions}

The asymptotical solutions derived analytically in this section are
confirmed by numerical integration of the system \re{4.3}--\re{4.5}.
Of course, numerical solutions can be tabulated in a finite interval
of evolution parameter. Nevertheless, a behavior of every numerical
solution considered at a large value of $\tau$ corresponds to one of
asymptotical solutions derived above. Hodographs of typical
numerical solutions are presented in figures
\ref{fig3}.1--\ref{fig3}.5, the monotonous solution ``1'',
eqs. \re{4.6}, \re{4.9}, exists for arbitrary parameters $\delta$
and $\chi$ (except $\delta=0$ and/or $\chi=0$), and $\delta$--$\chi$
domains in which other solutions ``2--5'' exist are outlined
in figure \ref{fig3}.D.
%
%---------------------- Figure 3 ----------------------------
\begin{figure*}
\begin{center}
\includegraphics[scale=0.7]{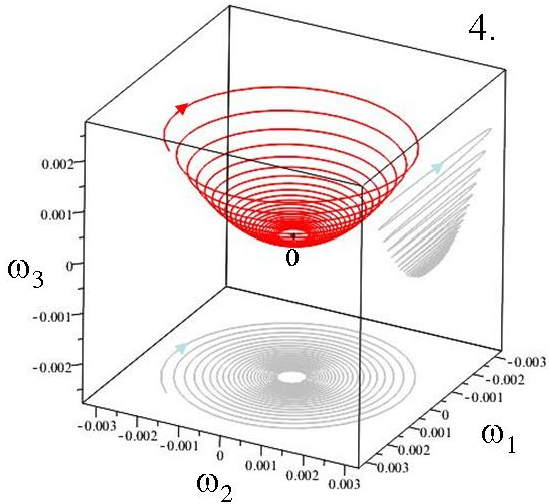}\hspace{5ex}
\includegraphics[scale=0.7]{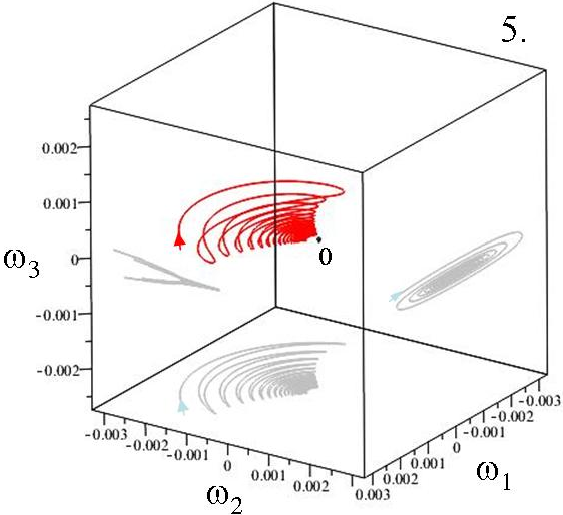}\vspace{5ex}\\
\includegraphics[scale=0.7]{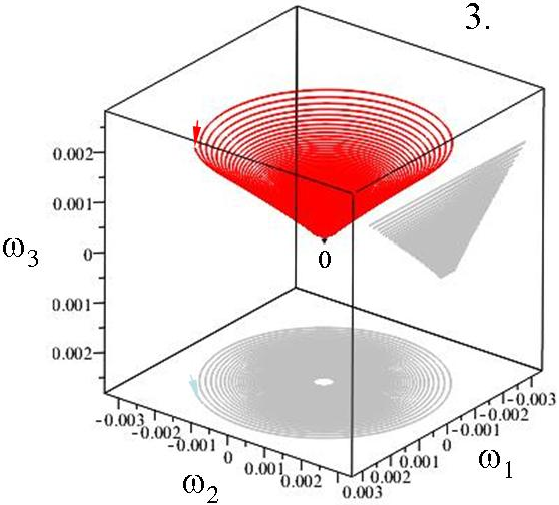}\hspace{5ex}
\includegraphics[scale=0.3]{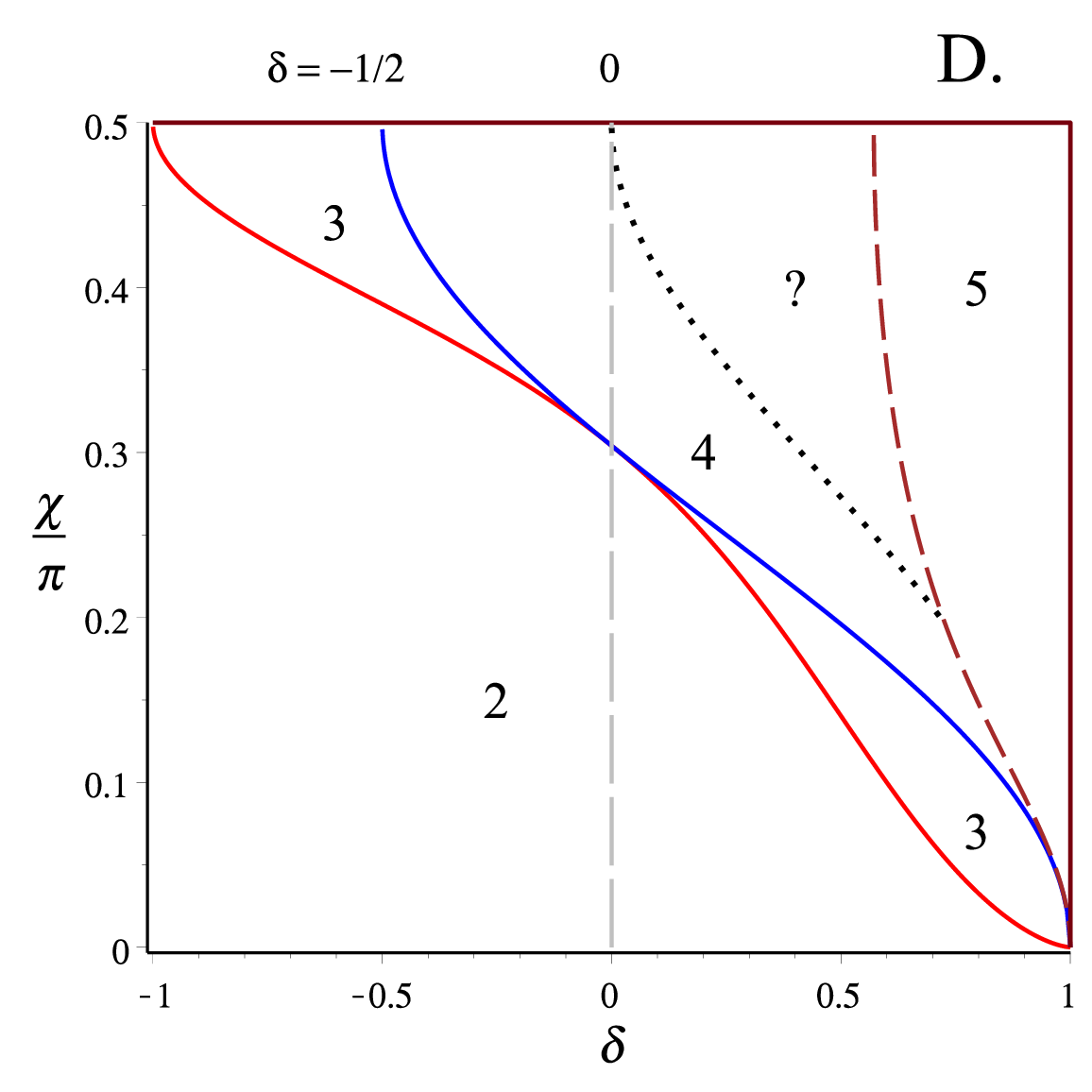}\vspace{5ex}\\
\includegraphics[scale=0.7]{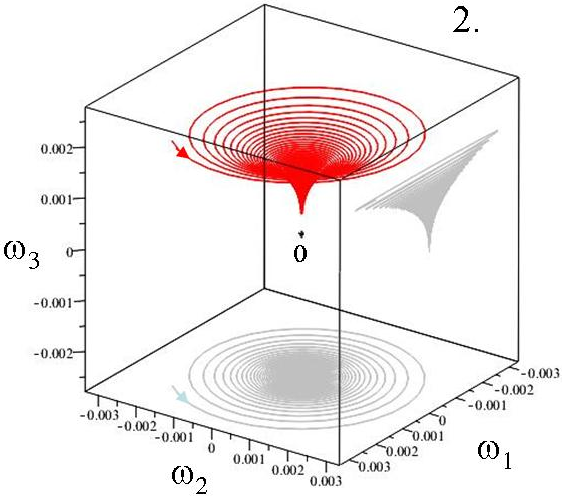}\hspace{5ex}
\includegraphics[scale=0.7]{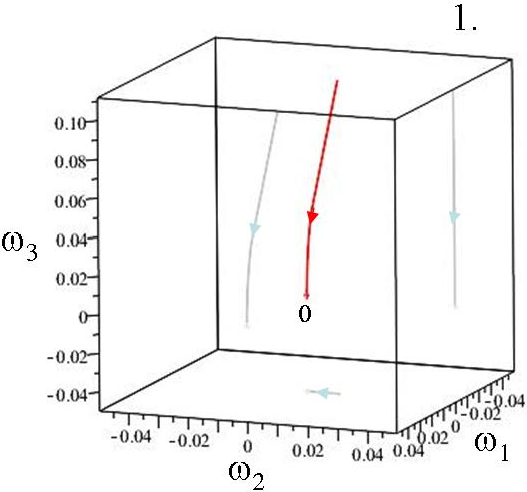}\vspace{5ex}
\caption{Hodographs of the dimensionless angular velocity $ \Bf\omega$
of the spinning top following numerical solutions with different
asymptotical behavior derived analytically.} \lab{fig3}
\end{center}
\end{figure*}
%------------------------------------------------------------
%
It is seen that the domain ``5'' of the corresponding oscillating
asymptotical solution overlaps the domain ``?'' partially (compare
figures \ref{fig2} and \ref{fig3}.D) but this is a consequence of
imperfection of the approximation used. Actually, numerical spiral
solutions ``4'' and oscillating solutions ``5'' (and also the
monotonous ones ``1'') coexist in the domains ``4'', ``?'' and ``5''
of the parameter diagram \ref{fig3}.D, and a specific evolution of
the spinning top in these domains depends on initial conditions.

We recall that aforementioned results are concerned with the cases
where a direction of the dipole moment vector does not coincide with
any principal axis of inertia. Thus the axial symmetry of the
problem is broken.

%%%%%%%%%%%%%%%%%%%%%%%%%%%% SECTION 5 %%%%%%%%%%%%%%%%%%%%%%%%%%%%%%%%%

\section{Special cases}
\renewcommand{\theequation}{5.\arabic{equation}}
\setcounter{equation}{0}

Here we consider three special cases where the system possesses a
symmetry: L) $\chi=0$, i.e., $ \Bf d\|  \Bf e_3$, the case of
longitudinal polarization; T) $\chi=\pi/2$, i.e., $ \Bf d\| \Bf e_1$,
the case of transversal polarization; O) $\delta=0$, i.e.,
$I_1=I_2=I_3$, the case of spherical top where the direction of $ \Bf d$ alone singles out the symmetry axis of the problem. These cases
were partially considered in \cite{Duv20a,Duv22a}. Here we outline
briefly and complete those results.

\subsection{Longitudinal polarization}

If $\chi=0$, i.e., $ \Bf d\| \Bf e_3$, the problem possesses the axial
symmetry. The dimensionless equations of motion \re{4.2}--\re{4.4}
take the form:
%
%           Equation 5.1
\begin{eqnarray}
\dot\omega_1-\delta\,\omega_2\omega_3&=& - \omega_a^2\omega_1,
\nn\\
\dot\omega_2+\delta\,\omega_1\omega_3&=&-\omega_a^2\omega_2,
\nn\\
(1-\delta)\dot\omega_3&=&0. \lab{5.1}
\end{eqnarray}
In view of the axial symmetry one can put
$\omega_1(0)\equiv\omega_{10}=0$. The set \re{5.1} possesses the
exact general solution:
%
%           Equation 5.2
\begin{eqnarray}
\omega_1&=&\omega_{20}\Phi\sin(\delta\omega_3\tau),\quad
\omega_2=\omega_{20}\Phi\cos(\delta\omega_3\tau),\nn\\
\omega_3&=&\omega_{30}=\,{\rm const},\quad \mbox{where}\quad \Phi=\sqrt{\frac{\kappa-1}{\kappa\mathrm{e}^{2\epsilon^2\omega_3^2\tau}-1}}\nn\\
\kappa&=&1+\epsilon^2\frac{\omega_3^2}{\omega_{20}^2},\quad\epsilon=1-\delta=\frac{I_3}{I_1}.
\lab{5.2}
\end{eqnarray}
Thus, $ \Bf\omega_\bot\to0$ in the limit $\tau\to\infty$, and the
hodograph of $ \Bf\omega$ spirales along the plane $\omega_3=$\,const
to the fixed point $ \Bf\omega\to \Bf\omega_\infty=\omega_3 \Bf e_3\ne0$
exponentially, by the characteristic time
$\Theta=1/(\tau_0\Omega_3^2I_3^2/I_1^2)$; see figure \ref{fig4}.L.

Quite different evolution of the longitudinally polarized symmetric
spinning top follows from the truncated Euler equation \re{2.5}
split by coordinate components \cite{Duv22a} and put here into the
dimensionless form:
%
%           Equation 5.3
\begin{eqnarray}
\dot\omega_1-\delta\,\omega_2\omega_3&=& - \omega_\bot^2\omega_1,
\nn\\
\dot\omega_2+\delta\,\omega_1\omega_3&=&-\omega_\bot^2\omega_2,
\nn\\
\dot\omega_3&=&-\omega_\bot^2\omega_3. \lab{5.3}
\end{eqnarray}
The solution (corresponding to $\omega_{10}=0$) \cite{Duv22a}
%
%           Equation 5.4
\begin{align}
\omega_1&=\frac{\omega_{20}}{F}\sin\!\left\{\frac{\delta\omega_{30}[F{-}1]}{\omega_{20}^2}\right\},\nn\\
\omega_2&=\frac{\omega_{20}}{F}\cos\!\left\{\frac{\delta\omega_{30}[F{-}1]}{\omega_{20}^2}\right\},
\nn\\
\omega_3&=\frac{\omega_{30}}{F},\quad \mbox{where}\ \
F\equiv\sqrt{1+2\omega_{20}^2\tau},   \lab{5.4}
\end{align}
describes the asymptotical power-law spiral decrease of the angular
velocity $ \Bf\omega\sim1/\sqrt{2\tau}\to0$ at $\tau\to\infty$ along
the cone surface, as on figure \ref{fig3}.3. Being natural at the
first glance, this solution carries non-physical consequences, as it
will be shown in the next section.

The solution \re{5.2} to the set \re{5.1} is valid also for the
spherical spinning top. In this case $\delta=0$, the precession of
the spinning top is absent, and the spiral hodograph degenerates to
the straight line on the plane $\omega_3=$\,const connecting the
initial point $ \Bf\omega_{0}$ and the final one
$ \Bf\omega_\infty=\omega_3 \Bf e_3$.

Respectively, the hodograph of the
solution \re{5.4} to the set \re{5.3} reduces at $\delta=0$ to the
straight line connecting $ \Bf\omega_{0}$ and $\Bf0$.

\subsection{Transversal polarization}

In the case $\chi=\pi/2$ the dimensionless equations of motion
\re{4.2}--\re{4.4} take the form:
%
%           Equation 5.5
\begin{eqnarray}
\dot\omega_1-\delta\,\omega_2\omega_3&=& 0,
\nn\\
\dot\omega_2+\delta\,\omega_1\omega_3&=&-\omega_c^2\omega_2,
\nn\\
(1-\delta)\dot\omega_3&=&-\omega_b^2\omega_3, \lab{5.5}
\end{eqnarray}
and an exact solution is still not known. Thus we apply to
approximations.

The spiral solutions ``3'' and ``4'', derived for
askew-polarized spinning top by the averaging method, are valid in
the present case. They reveal the cone-shape trajectories (figure
\ref{fig3}.3) if $-1\ge\delta\ge-1/2$, and the bowl-shape
trajectories (figure \ref{fig3}.4) if $-1/2<\delta<0$. For the
prolate spinning top $\delta>0$ we should apply to the oscillating
solution ``5'', but it does not exist in the case
$\chi=\pi/2$. Indeed, since $n_3=0$, the magnitude $M=0$, and then
$C=0$, as it follows from \re{4.42}. Hence the phase variable
$\varphi$ ``stops'': $\varphi=0$, and the averaging method is not
applicable in this case. Thus the behavior of the prolate
transversally polarized spinning top needs more study.

In contrast to the general case, in the case $\chi=\pi/2$ the dipole
vector $ \Bf d$ can be directed along the inertia axis $0 \Bf e_1$ which
is the rotation symmetry axis of order 2. Thus the system of
equations \re{5.3} possesses the set of fixed points
$\omega_\infty \Bf e_1$, $\omega_\infty\in{\Bbb R}$ covering this
inertia axis. Given $\omega_\infty$, one can linearize the system
\re{5.5} in the vicinity $\omega_1\approx\omega_\infty$,
$\vl\omega_2\vl, \vl\omega_3\vl\ll\vl\omega_\infty\vl$ of the fixed point
$ \Bf\omega_\infty=\omega_\infty \Bf e_1$:
%
%           Equation 5.6
\begin{eqnarray}
\dot\omega_1&=&0,
\nn\\
\dot\omega_2&=&-\omega_\infty^2\omega_2-\delta\omega_\infty\omega_3,
\nn\\
\dot\omega_3&=&-\frac{1+2\delta}{1-\delta}\,\omega_\infty^2\omega_3.
\lab{5.6}
\end{eqnarray}
A search of exponential solutions $ \Bf\omega-\omega_\infty \Bf e_1\sim\,{\rm e}^{\lambda\tau}$ leads to the characteristic numbers
%
%           Equation
\begin{equation*}
\lambda_1=0,\quad  \lambda_2=-\omega_\infty^2,\quad
\lambda_3=-\frac{1+2\delta}{1-\delta}\,\omega_\infty^2
\end{equation*}
and to the corresponding solution of Cauchy problem
%
%           Equation 5.7
\begin{eqnarray}
\omega_1&=&\omega_\infty,
\nn\\
\omega_2&=&\omega_{20}\,{\rm e}^{\lambda_2\tau}
+\frac{1-\delta}{3}\frac{\omega_{30}}{\omega_\infty}\left(\,{\rm
e}^{\lambda_3\tau}-\,{\rm e}^{\lambda_2\tau}\right),
\nn\\
\omega_3&=&\omega_{30}\,{\rm e}^{\lambda_3\tau}. \lab{5.7}
\end{eqnarray}
%
%
%
%---------------------- Figure 4 ----------------------------
\begin{figure}
\begin{center}
\includegraphics[scale=0.6]{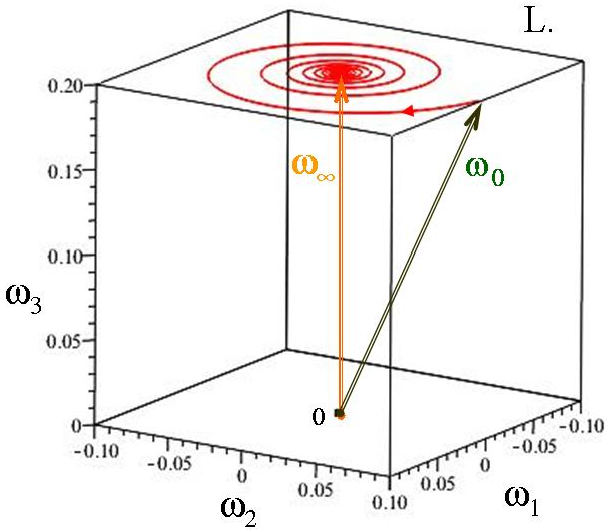}\bigskip\\
\includegraphics[scale=0.6]{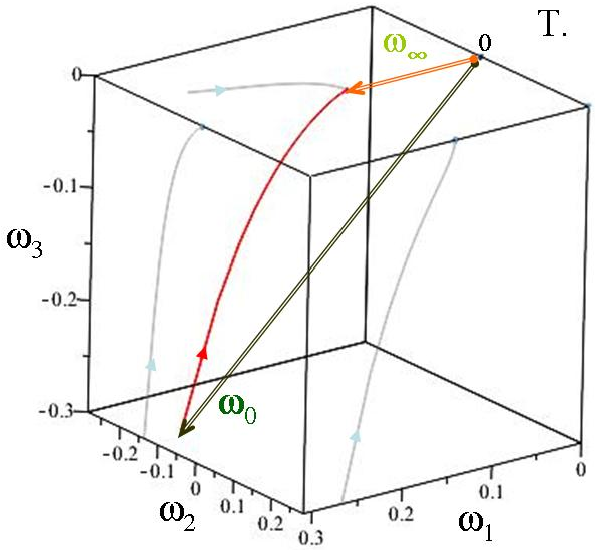}
\caption{The spinning top in special cases.
L.\,Longitudinal polarization: exact solution. T.\,Transversal
polarization: numerical solution. In both cases the dimensionless
angular velocity tends asymptotically to its residual value
$ \Bf\omega_\infty$ directed along the dipole vector $ \Bf d$.\lab{fig4}}
\end{center}
\end{figure}
%------------------------------------------------------------
This solution is stable, i.e., $\omega_2\to0$, $\omega_3\to0$,
provided $\lambda_3<0$. The latter condition holds for
$-1/2<\delta<1$. In the domain $-1/2<\delta<0$ the solution \re{5.7}
overlaps with the spiral asymptotical solution ``4''. Thus two
scenarios are possible: the system reaches exponentially the
non-zero fixed point $ \Bf\omega_\infty=\omega_\infty \Bf e_1$, if the
initial value $\vl\, \Bf\omega_\bot(0)\vl\ll\omega_\infty$, or the sinning
top slows down to stop by the power law
$\omega\sim1/\sqrt{\tau}\to0$. Both scenarios are approved by
numerical calculations \cite{Duv22a}; compare figures \ref{fig3}.4
and \ref{fig4}.T.

%%%%%%%%%%%%%%%%%%%%%%%%%%%% SECTION 6 %%%%%%%%%%%%%%%%%%%%%%%%%%%%%%%%%

\section{Two examples}
\renewcommand{\theequation}{6.\arabic{equation}}
\setcounter{equation}{0}

\subsection{The cellulose nanocrystal}

The example of an axially-symmetric composite particle with a large
longitudinal dipole moment is a cellulose nanocrystal of the slender
rod shape with the following characteristics: length $L=315\,$nm,
its ratio to diameter $L/D=10$, density $\rho=1.6\,$g/cm$^3$. The
dipole moment $d=4400\,$Debye is directed along a symmetry axis
\cite{F-PBL14}. Thus the relevant to this case solution of the reduced Euler
equations is given by eqs. \re{5.2}. It can be easily expressed in terms
of the original dimensional variables by means of the relations \re{4.1}.

Approximating the nanocrystal shape by an elongated cylinder yields
the following principal inertia moments:
$I_1=3.3\cdot10^{-26}\,$g$\cdot$cm$^2$,
$I_3=4.11\cdot10^{-28}\,$g$\cdot$cm$^2$. Consequently, the
characteristic time scale parameter \re{3.8} is
$\tau_0=1.4\cdot10^{-35}\,$c, and the ratio $\epsilon=I_3/I_1=0.015$
is a small parameter. Given the value of $\Omega_3=\Omega_{30}$ even
record, $\Omega_3\sim\pi\cdot10^{10}\,$c$^{-1}$, the exponential drift
$\BOm_\bot\to0$ is very slow, with the astronomical braking time
$\Theta=1/(\epsilon^2\tau_0\Omega_3^2)\sim10^{10}\,$years. Thus,
this asymptotical regime is unattainable and can be pushed to
infinity by taking the limit $\epsilon\to0$ in the solution
\re{5.2}, \re{4.1}:
%
%           Equation 6.1
\begin{eqnarray}
\Omega_1&\mathop{\longrightarrow}\limits_{\epsilon\to0}&\frac{\Omega_{20}}{F(t)}\sin(\Omega_3t),\quad
\Omega_2\mathop{\longrightarrow}\limits_{\epsilon\to0}\frac{\Omega_{20}}{F(t)}\cos(\Omega_3t),\nn\\
\Omega_3&=&\Omega_{30}, \lab{6.1}
\end{eqnarray}
where the square root function $F(t)$ is defined by \re{3.9}. This
power-law decrease of $\BOm_\bot\to0$ is most intensive during the
time $T=1/(\tau_0\Omega_{20}^2)\sim3\cdot10^6\,$years which,
although not astronomical, is nevertheless geological, i.e., too
large for laboratory observations.

Following the truncated Euler equations \re{5.3}, the limit
$\epsilon\to0$ of the corresponding solution \re{5.4} is similarly
power-law
%
%           Equation 6.2-3
\begin{align}
\Omega_1\mathop{\longrightarrow}\limits_{\epsilon\to0}&\frac{\Omega_{20}}{F(t)}\sin\Omega_{30}\vartheta(t),\nn\\
\Omega_2\mathop{\longrightarrow}\limits_{\epsilon\to0}&\frac{\Omega_{20}}{F(t)}\cos\Omega_{30}\vartheta(t),\quad
\Omega_3\mathop{\longrightarrow}\limits_{\epsilon\to0}\frac{\Omega_{30}}{F(t)}, \lab{6.2}\\
\mbox{where}&\quad
\vartheta(t)\equiv\frac{F(t)-1}{\tau_0\Omega_{20}^2},
\lab{6.3}
\end{align}
but here all three components of the angular velocity decrease with
the same geological characteristic time
$T=1/(\tau_0\Omega_{20}^2)\sim3\cdot10^6\,$years.

Actually, the difference between the solutions \re{6.1} and \re{6.2}
is minor. This is seen better by studying a motion of the
nanocrystal in space.

To have a complete description of the rotary dynamics of the rigid
body in space, the Euler equations should be complemented by the
Poisson equations relating the components of the angular velocity
$\BOm=\{\Omega_1,\Omega_2,\Omega_3\}$ as functions of $t$ with the
Euler angles $\varphi, \theta, \psi$ and their time derivatives.
This set of differential equations has the following normal form
\cite{Duv22a}:
%
%           Equation 6.4
\begin{eqnarray}
\dot\varphi &=& (\Omega_1\sin\psi + \Omega_2\cos\psi)/\sin\theta,
\nn\\
\dot\theta &=& \Omega_1\cos\psi - \Omega_2\sin\psi, \nn\\
\dot\psi  &=& \Omega_3 - (\Omega_1\sin\psi +
\Omega_2\cos\psi)\cot\theta. \lab{6.4}
\end{eqnarray}

Since all initial orientations of the free rigid body in an
isotropic space are physically equivalent, it is sufficient to find
any partial solution of the equations \re{6.4}. Substituting
\re{6.1} into \re{6.4} yields the exact solution:
%
%           Equation 6.5
\begin{align}
\varphi(t)=\Omega_{20}\vartheta(t), \quad \theta=\frac\pi2, \quad
\psi(t)=\Omega_3t.& \lab{6.5}
\end{align}
It describes a composition of the slowing flatwise rotation of the
cellulose rod and its uniform proper rotation around the symmetry
axis with the constant angular velocity $\Omega_3$.

The functions \re{6.2}--\re{6.3} yield a similar to \re{6.5}
solution of the Poisson equations:
%
%           Equation 6.6
\begin{align}
\varphi(t)&=\Omega_{20}\vartheta(t), \ \ \theta=\frac\pi2, \ \
\psi(t)=\Omega_{30}\vartheta(t), \lab{6.6}
\end{align}
with that difference the proper rotation is not uniform. But the
rotation of the rod around its symmetry axis is hardly observable in
the $D/L\to0$ limit of an infinitely thin rod.

In all respects the cellulose nanocrystals possessing even enormous
dipole moment looks to be poor objects for the experimental
verification of the radiative spindown.

\subsection{The organic DAST-nanocrystal}

The example of the permanently
askew-polarized composite particle might be the hypothetical organic
polar monocrystal of
4-dimethylamino-N-methyl-4-stilbazolium tosylate
(DAST), chemical formula C$_{23}$H$_{26}$N$_{2}$SO$_{3}$, which is
not synthesized yet but its properties are predicted in literature;
see \cite{MNS03} and refs. therein. Crystallographic axes are almost
orthogonal thus, if a monocrystal is naturally confined by
crystallographic planes, its natural form is a parallelepiped. The
dipole moment of a single ion pair is 30\,Debyes, it lies in the
plane 0$ \Bf e_1 \Bf e_3$ and is inclined to the axis  0$ \Bf e_1$
by the angle $35^\circ$. Taking into account the lattice parameters
of such crystal: the cell volume
$v=1.04$\,nm$\times1.13$\,nm$\times1.79$\,nm$=2.1$\,nm$^3$, the
molecular weigh $\mu=410.5$, the number of molecules per cell $Z=4$,
one can calculate the density $\rho=1.37$\,g/cm$^3$ and then the
principal inertia moments of the nanocrystal. Moreover, the dipole
moment of DAST-nanocrystals is predicted to behave as an additive
vector characteristic. Thus one can imagine the DAST-nanocrystal of
an arbitrary dimension and a shape, and calculate its inertia tensor
and the dipole moment vector.

We are interested in axially-symmetric nanocrystals. Two families
are possible.

%
%---------------------- Figure 5 ----------------------------
\begin{figure*}[tb]
\begin{center}
\includegraphics[scale=0.52]{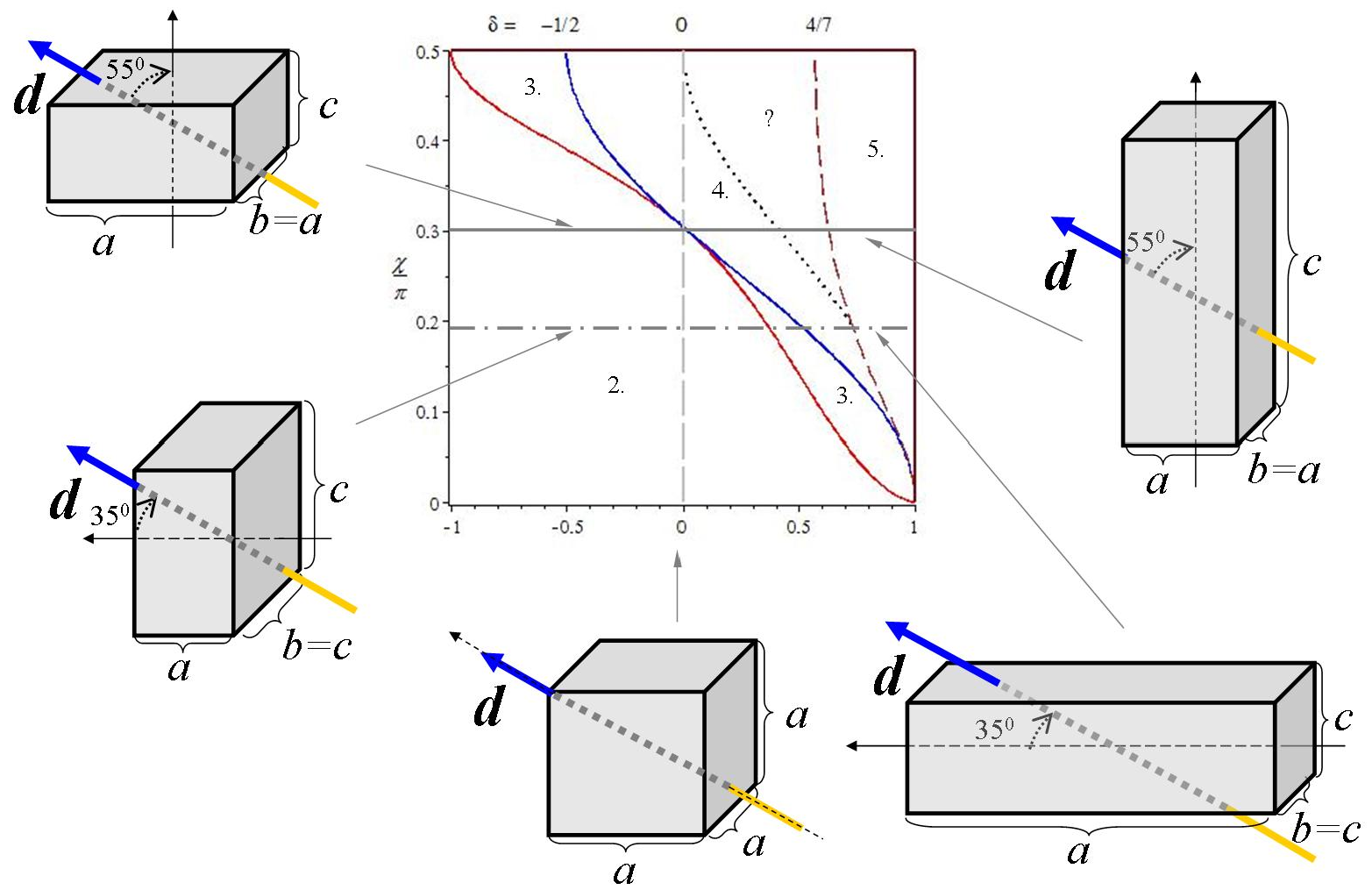}
\vspace{0ex} \caption{Two families of axially-symmetric
DAST-monocrystals. The 1st family consists of monocrystals of the
dimensions $a=b\ne c$ with the dipole moment $ \Bf d$ inclined to the
symmetry axis 0$ \Bf e_3$ by the angle $55^\circ=0.306\pi$ (the solid
gray line in the $\delta$--$\chi$ diagram). The 2nd family
corresponds to the dimensions $a\ne b=c$ with the dipole moment
inclined to the symmetry axis 0$ \Bf e_1$ by the angle
$35^\circ=0.194\pi$ (the dash-dotted gray line). In the $\delta=0$
case of cubic crystals one of principal inertia axes can be directed
along $ \Bf d$. \lab{fig5}}
\end{center}
\end{figure*}
%------------------------------------------------------------

The first family consists of monocrystals of the dimension $a=b\ne
c$ with the dipole moment inclined to the symmetry axis 0$ \Bf e_3$ by
the angle $\chi=55^\circ$. The rate of sides $c/a$ determines the
rate $I_3/I_1$ of the principal inertia moments $I_1=I_2\ne I_3$,
and thus the elongacy parameter
$\delta=1-I_3/I_1=(c^2-a^2)/(c^2+a^2)$. The DAST-nanocrystal of the
dimensions $a=b=100$\,nm, $c=50$\,nm with inertia moments
$I_1=I_2=7.14\cdot10^{-27}$\,g$\cdot$cm$^2$, $I_3/I_1=1.6$ (so that
$\delta=-0.6$) may possess a huge dipole moment $d=2.8\cdot10^7$\,D
\cite{MNS03}. Varying the dimensions $a=b$ and $c$ changes the
inertia moments $I_1$, $I_3$ and the total dipole moment $d$ but
keeps unchanged the inclination angle $\chi=55^\circ$. Thus the only
free dimensionless parameter determining the dynamical system
remains, i.e., the inertial characteristic $\delta$.

The line corresponding to the 1st family of DAST-monocrystals on the
$\delta$--$\chi$ diagram (figure \ref{fig5}) crosses the domain
``3'' close to its isthmus at $\delta=0$,
$\chi_0=0.304\pi=54.7^\circ\approx55^\circ$, thus the corresponding
solution does not apply to this family.

The second family of axially-symmetric DAST-monocrystals corresponds
to the dimensions $b=c\ne a$ with the dipole moment inclined to the
symmetry axis 0$ \Bf e_1$ by the angle $35^\circ$. This family admits
all types of solutions presented on figure \ref{fig3}.

Special interest should be paid to the cubic DAST-monocrystals which
are equivalent, by inertial properties, to the spherical spinning
top. In despite the dipole moment is still inclined to
crystallographic axes, the principal inertia exes can be oriented
arbitrarily. Choosing 0$ \Bf e_3\|  \Bf d$ reduces the dynamical problem
to the solvable $\delta=0$ case considered at the end of subsection
5.1.

Substituting $\delta=0$ and, consequently, $\epsilon=1$ into eqs.
\re{5.2}, \re{4.1} yields the following solution of the reduced
Euler equations:
%
%           Equation 6.7
\begin{eqnarray}\lab{6.7}
\Omega_1&=&0,\quad
\Omega_2=\Omega_{20}\vl\Omega_3\vl\left(\Omega_0^2\mathrm{e}^{2\tau_0\Omega_3^2t}-\Omega_{20}^2\right)^{-1/2},\nn\\
\Omega_3&=&\Omega_{30}.
\end{eqnarray}
The corresponding exact solution of the Poisson equations \re{6.4}
is unknown. One can find the asymptotic dependency of the Euler
angles on time:
%
%           Equation 6.8
\begin{align}
\psi&\sim0, \quad \theta(t)\sim\mathrm{e}^{-\tau_0\Omega_3^2t},
\quad \varphi(t)\sim\Omega_3t. \lab{6.8}
\end{align}
Hence, at $t\to\infty$, the DAST crystal rotates around the vertical
axis with the angular velocity $\sim\Omega_3$ in such a way that the
dipole vector ${ \Bf{\frak d}}$ spirales to this axis exponentially,
with the characteristic time $\Theta=1/(\tau_0\Omega_3^2)$.

The solution of the truncated Euler equations follows from eqs.
\re{5.4}:
%
%           Equation 6.9
\begin{eqnarray}\lab{6.9}
\Omega_1&=&0,\quad \Omega_2=\frac{\Omega_{20}}{F(t)},\quad
\Omega_3=\frac{\Omega_{30}}{F(t)},
\end{eqnarray}
and the corresponding exact solution o the Poisson equations
\re{6.4} is:
%
%           Equation 6.10
\begin{align}
\psi&=0, \ \ \theta=\arctan\,\rule[-2ex]{0.5pt}{5.6ex}\,\frac{\Omega_{20}}{\Omega_{30}}\,\rule[-2ex]{0.5pt}{5.6ex}\, , \ \
\varphi(t)=\mathrm{sign}\Omega_3\,\Omega_0\vartheta(t); \lab{6.10}
\end{align}
here the power-law functions $F(t)$ and $\vartheta(t)$ are defined
in \re{3.9} and \re{6.3}, respectively. This solution predicts the
slowing rotation of the crystal to stop around the vertical axis
with the dipole vector ${ \Bf{\frak d}}$ inclined to this axis by the
constant angle $\theta$.

For the cubic DAST-nanocrystal of the dimension $a=100\,$nm with
equal inertia moments $I=2.3\cdot10^{-26}$\,g$\cdot$cm$^2$ and the
dipole moment $d=5.6\cdot10^7$\,D we have the time scale
$\tau_0=3.4\cdot10^{-27}\,$c. Thus, if the initial angular velocity
$\Omega_{30}\sim\Omega_{20}\sim\Omega_0\sim\pi\cdot10^{10}\,$Hz, as in
the experiment \cite{JYRLYZ21}, the characteristic braking time
$\Theta\sim1/(\tau_0\Omega_0^2)\sim4\,$days which is of order of a
storage time in Penning trap \cite{Haf03}. Then such a nanocrystal
might serve as a test for choosing between the reduced Euler
equation, and the truncated one, that is between the balance
equations \re{2.1} and \re{2.4a}.

%\newpage
%%%%%%%%%%%%%%%%%%%%%%%%%%%% SECTION 6 %%%%%%%%%%%%%%%%%%%%%%%%%%%%%%%%%

\section{Conclusions}

In the previous paper \cite{Duv20a} the equation of rotary motion of
a rigid body with the permanent electric dipole moment \re{2.2} was
derived from the Landau-Lifshitz angular momentum balance condition
\cite[section 75]{L-L87E}. This singularly perturbed equation of the 2nd
order with respect to the angular velocity $\BOm$ is then reduced to
the 1st-order regularly perturbed equation by means of the
unperturbed Euler equation and its differential consequence
\re{2.3}.

Another common way to get rid of higher derivatives is to abandon
the Schott-type term in the expression for the the radiation
reaction torque. It leads to the truncated Euler equation, the
correctness of which raises doubts \cite{Duv22a}.

Here the the case of askew-polarized spinning top with
axially-symmetric inertia ellipsoid is considered. This case is
suggested by the theoretical works predicted a giant dipole moment
of the DAST-nanocrystal; see \cite{MNS03} and refs. therein. The
dipole moment vector which is inclined to the symmetry axis spoils
the axial symmetry of the problem and complicates the dynamics, when
compared to the symmetric case \cite{Duv20a,Duv22a}.

In general case of inclined dipole moment by the angle
$0<\chi<\pi/2$ a general exact solution of the reduced Euler-type
equations \re{3.3}--\re{3.5} is unknown. The only partial exact
solution \re{3.9} has been found. Actually, it reveals some general
features of the spinning top evolution studied analytically and
numerically. Given the initial velocity $\Omega_0=\vl\BOm(0)\vl$, the
spinning top slows its rotation down most intensively during the
characteristic time $T\sim1/(\tau_0\Omega_0^2)$. Then the top
gradually switches up to the asymptotic regime in which, at $t\gg
T$, the magnitude $\Omega$ of the angular velocity decreases by the
power law $\Omega\sim1/\sqrt{\tau_0t}$, independently on the initial
values.

Details of the asymptotical evolution depend mainly on parameters of
the polarized spinning top, its elongacy $\delta$ and the
inclination angle $\chi$, and less on initial data.

First of all we recast the equations of motion in the dimensionless
form \re{4.2}--\re{4.5} and used asymptotic expansions to examine a
monotonous power-law behavior \re{4.6} of the angular velocity
components at $\tau\to\infty$. We arrive at the asymptotic solution
\re{4.6}, \re{4.9}. Numerical calculations show that this
asymptotics is inherent to a rather narrow set of initial data.

Then, regarding the radiation reaction torque as a small
perturbation (what is actually true), we considered the precessing
rotation of the symmetric top and studied its secular evolution
using the averaging method.

The averaging method appears not applicable at some values of
parameters $\delta$ and $\chi$ (the domain ``?'' in figure
\ref{fig2}), thus we considered the perturbation of the exact
solution \re{3.9}. This perturbed solution makes a sense on the
finite time interval, but it has been extended up to a time infinity
by the combination of the asymptotic expansion, the averaging and
the linearization methods.

As a result, we found five general types of the askew-polarized
spinning top evolution which differ essentially from one another but
reveal the common asymptotical power law
$\Omega\sim1/\sqrt{\tau_0t}$ of a spin down.

The monotonous solution ``1'' describes a decelerating
rotation mainly around the symmetry axis $0 \Bf e_3$ with quickly
ceasing rotations in other directions. This solution is possible for
any parameters $-1<\delta<1$, $0<\chi<\pi/2$ if the hodograph of
angular velocity passes sufficiently close to the point determined
by eqs. \re{4.6}, \re{4.9} in some instant $t$. Otherwise, the top
undergoes a slowly varying precession. Respectively, the hodograph
spirals to zero along a surface which shape depends on parameters
$\delta$ and $\chi$ of the top: the funnel-shape surface in the
domain ``2'', the bowl-shape surface in the domain
``4'', and the cone in the domain ``3''; see figures
\ref{fig2} and \ref{fig3}.2--4. Respectively, the instant rotation
axis tends asymptotically to the symmetry axis in the domain
``2'' and to the orthogonal plane in the domain ``4''
while in the domain ``3'' the precessing rotation slows down
at a constant nutation angle.

In the domain ``?'' the rotation of the spinning top cannot
be considered as a slowly varying precession around the symmetry
axis $0 \Bf e_3$. Thus we apply to the exact solution \re{3.9}
describing a rotation of the top around the axis $0 \Bf e_2$ and
consider a perturbation to this solution. It appears unstable but
suggests the asymptotical solution ``5'' with wedge-shape
hodograph. The vector of angular velocity ceases in this case by
making slow ample waggles around the symmetry axis and narrow
waggles in the orthogonal plane. This motion is characteristic for
markedly prolate top with sufficiently inclined dipole moment.

In all these cases concerning the askew-polarized spinning top the
angular velocity decreases asymptotically to zero by the power law
$\Omega\sim1/\sqrt{\tau_0t}$.

Another behavior can be revealed by the spinning top with
longitudinal polarization. In this case the reduced Euler equations
possess non-zero fixed points and admit an exponential drift of the
angular velocity to these points in the characteristic time
$\Theta\sim1/(\tau_0\Omega_{\|}^2I_\|^2/I_\bot^2)$ depending on the
conserved angular velocity component $\Omega_\|$ along the symmetry
axis. Similar behavior may occur in the case of transversal
polarization \cite{Duv20a}. As a result, the spinning top tends to
the final state in which it rotates forever with some residual
angular velocity around the dipole moment vector.

On the contrary, the truncated Euler equations predict a
power-law spindown for even symmetric particles with the
characteristic braking time $T\sim1/(\tau_0\Omega_{\bot0}^2)$
depending on the initial transversal angular velocity
$\Omega_{\bot0}$.

As an experimental test for the correct radiation reaction torque
may serve strongly polarized nanoparticles. But even enormous dipole
moment 4400\,D of the existing slender cellulose nanocrystals
\cite{F-PBL14} is not sufficient to detect the radiation reaction
spindown. Indeed, being spun up to GHz they take cosmological period
of $\Theta\sim10^{10}\,$years to go to the final state of residual
rotation predicted by the reduced Euler equations. Instead, during
the much shorter but geological time of $T\sim10^{6}\,$years one
might observe a spindown by the power law which coincides with that
predicted by the truncated equations.

A better experimental test to distinguish between two aforementioned
frameworks might be a cubic DAST-nanocrystal of size, say
$a=100$\,nm, possessing, in theory \cite{MNS03}, the giant dipole
moment $d=5.6\cdot10^7$\,D . We suppose that it can be spun up to
the initial angular velocity $\Omega_0=\pi\cdot10^{10}\,$Hz, as in
the experiment \cite{JYRLYZ21}. Then such a nanocrystal could serve
as a test for choosing between balance equations \re{2.1} and
\re{2.4a}.

Following \re{2.1}, the nanocrystal changes markedly its rotation in
the characteristic time $T\sim1/(\tau_0\Omega_0^2)\sim4\,$days.
Namely, it drifts to its final state of a constant rotation around
the dipole vector. Following \re{2.4a}, the slowdown is most intense
during the same time and then it continues to stop by the power law
$\Omega\sim1/\sqrt{\tau_0t}$. The experimental problem in this test
is a storage of neutral polarized nanoparticles free or, at least,
almost free for few days. For this purpose might serve the currently
designing trap for polar particles \cite{PMY20}.

%%===========================================================================================%%
%% If you are submitting to one of the Nature Portfolio journals, using the eJP submission   %%
%% system, please include the references within the manuscript file itself. You may do this  %%
%% by copying the reference list from your .bbl file, paste it into the main manuscript .tex %%
%% file, and delete the associated \verb+\bibliography+ commands.                            %%
%%===========================================================================================%%

%\input{DAST_EPJD-L}

%\bibliography{sn-bibliography}% common bib file
%% if required, the content of .bbl file can be included here once bbl is generated
%%\input sn-article.bbl

%% BioMed_Central_Bib_Style_v1.01

%\end{document}

%%

\end{document}